%
\let\useblackboard=\iftrue
%
%
\newfam\black
\input harvmac.tex
\def\Title#1#2{\rightline{#1}
\ifx\answ\bigans\nopagenumbers\pageno0\vskip1in%
\baselineskip 15pt plus 1pt minus 1pt
\else
\def\listrefs{\footatend\vskip 1in\immediate\closeout\rfile\writestoppt
\baselineskip=14pt\centerline{{\bf References}}\bigskip{\frenchspacing%
\parindent=20pt\escapechar=` \input
refs.tmp\vfill\eject}\nonfrenchspacing}
\pageno1\vskip.8in\fi \centerline{\titlefont #2}\vskip .5in}

\ifx\answ\bigans\def\tcbreak#1{}\else\def\tcbreak#1{\cr&{#1}}\fi
\useblackboard
\message{If you do not have msbm (blackboard bold) fonts,}
\message{change the option at the top of the tex file.}
\font\blackboard=msbm10 scaled \magstep1
\font\blackboards=msbm7
\font\blackboardss=msbm5
\textfont\black=\blackboard
\scriptfont\black=\blackboards
\scriptscriptfont\black=\blackboardss

\else

\fi
%
\def\yboxit#1#2{\vbox{\hrule height #1 \hbox{\vrule width #1
\vbox{#2}\vrule width #1 }\hrule height #1 }}
\def\fillbox#1{\hbox to #1{\vbox to #1{\vfil}\hfil}}
\def\ybox{{\lower 1.3pt \yboxit{0.4pt}{\fillbox{8pt}}\hskip-0.2pt}}
\def\np#1#2#3{Nucl. Phys. {\bf B#1} (#2) #3}
\def\pl#1#2#3{Phys. Lett. {\bf #1B} (#2) #3}

\def\comments#1{}

\def\half{{1\over 2}}
\def\Tr{{{\rm Tr\  }}}
\def\tr{{\rm tr\ }}

\def\CF{{\cal F}}

\def\CL{{\cal L}}

\def\II{\relax{I\kern-.07em I}}

\def\IZ{\relax\ifmmode\mathchoice
{\hbox{\cmss Z\kern-.4em Z}}{\hbox{\cmss Z\kern-.4em Z}}
{\lower.9pt\hbox{\cmsss Z\kern-.4em Z}}
{\lower1.2pt\hbox{\cmsss Z\kern-.4em Z}}\else{\cmss Z\kern-.4em
Z}\fi}
\def\IB{\relax{\rm I\kern-.18em B}}
\def\IC{{\relax\hbox{$\inbar\kern-.3em{\rm C}$}}}
\def\ID{\relax{\rm I\kern-.18em D}}
\def\IE{\relax{\rm I\kern-.18em E}}
\def\IF{\relax{\rm I\kern-.18em F}}
\def\IG{\relax\hbox{$\inbar\kern-.3em{\rm G}$}}
\def\IGa{\relax\hbox{${\rm I}\kern-.18em\Gamma$}}
\def\IH{\relax{\rm I\kern-.18em H}}
\def\II{\relax{\rm I\kern-.18em I}}
\def\IK{\relax{\rm I\kern-.18em K}}
\def\IP{\relax{\rm I\kern-.18em P}}

\font\cmss=cmss10 \font\cmsss=cmss10 at 7pt
\def\IR{\relax{\rm I\kern-.18em R}}

\def\tilde{\widetilde}
\Title{ \vbox{\baselineskip12pt\hbox{hep-th/9612157}
\hbox{RU-96-117}}}
{\vbox{\centerline{Branes from Matrices}}}
\centerline{Tom Banks, Nathan Seiberg and Stephen Shenker}
\smallskip
\smallskip
\centerline{Department of Physics and Astronomy}
\centerline{Rutgers University }
\centerline{Piscataway, NJ 08855-0849}
\centerline{\tt banks, seiberg, shenker@physics.rutgers.edu}
\bigskip
\bigskip
\noindent
Various aspects of branes in the recently proposed matrix model for M
theory are discussed.  A careful analysis of the supersymmetry algebra
of the matrix model uncovers some central charges which can be
activated only in the large $N$ limit.  We identify the states with
non-zero charges as branes of different dimensions.

\Date{December 1996}

\newsec{\bf Introduction}

\nref\bfss{T. Banks, W. Fischler, S.H. Shenker and L. Susskind,
hep-th/9610043.}%
\nref\bd{M. Berkooz and M.R.Douglas, hep-th/9610236.}%
\nref\compact{L. Susskind, hep-th/9611164.}%
\nref\princeton{O. Ganor, S. Rangoolam and W. Taylor, hep-th/9611202.}%
\nref\ab{O. Aharony and M. Berkooz, hep-th/9611215.}%
\nref\mrd{M. R. Douglas, hep-th/9612126.}%

The matrix model approach to M theory \bfss\ has successfully passed a
number of consistency tests \refs{\bfss - \mrd}.  One outstanding
problem that remains to be solved is the proper description of five
branes in the matrix model.  Berkooz and Douglas \bd\ have introduced
five branes wrapped around the longitudinal direction of the light cone
frame by adding degrees of freedom to the original model of \bfss.  The
purpose of the present paper is to define a general formalism for the
study of Bogolmonyi Prasad Sommerfield (BPS) $p$-branes in the matrix
model.

The eleven dimensional supersymmetry (SUSY) algebra
\eqn\susyalg{\{ Q_\alpha, Q_\beta \} = 2P^\mu \gamma_{\mu\alpha\beta} +
2Z^{\mu_1\mu_2}\gamma_{\mu_1\mu_2\alpha\beta}+
2Z^{\mu_1...\mu_5}\gamma_{\mu_1...\mu_5\alpha\beta} } 
$(\mu_i = 0,...,10$; $\alpha,\beta = 1,...32$) includes two central
charges: a two brane charge $Z^{\mu_1\mu_2}$ and a five brane charge
$Z^{\mu_1...\mu_5}$.
These charges are Lorentz tensors and as such their existence seems to
violate the Coleman-Mandula theorem.  A contradiction is avoided by
noticing that when these charges do not vanish they are infinite.  This
infinity has a simple physical interpretation.  The charged objects are
branes which in flat eleven dimensional space have infinite volume.  The
total charge is infinite but the charge per unit volume is finite.  A
simple way of handling these infinities is to assume that space is
compactified on a very large but finite torus.  Then, the charges are
all finite.  Since we keep the time non-compact, $Z^{0\mu_2} =
Z^{0\mu_2...\mu_5}=0$.  Equivalently, this result follows from the fact
that the Lorentz indices of the conserved currents associated with these
central charges $j_{\mu_1\mu_2\mu_3}$ and $j_{\mu_1...\mu_6}$ are
antisymmetric. The charges are given by integrals over space of
$j_{\mu_1\mu_ 20}$ and $j_{\mu_1...0}$ and therefore $Z^{0\mu_2} =
Z^{0\mu_2...\mu_5}=0$.

In the light cone frame only $SO(9) \subset SO(10,1)$ is manifest -- the
11 coordinates $x^\mu$ become $x^\pm, x^a$ ($a=1...9$) and the 32
supercharges become a pair of 16 supercharges $Q_\alpha$ ($\alpha =
1...16$) which we will refer to as dynamical supercharges and $\tilde
Q_\alpha$ which are kinematical.  The supersymmetry algebra is
\eqn\lcSUSY{\eqalign{
&\{\tilde Q_\alpha, \tilde Q_\beta \}= 2P^+\delta_{\alpha \beta} \cr
&\{Q_\alpha, \tilde Q_\beta \}= 2P^a \gamma_{a\alpha \beta} +
2Z^{a_1a_2}\gamma_{a_1a_2\alpha \beta}+
2Z^{a_1...a_5}\gamma_{a_1...a_5\alpha \beta} \cr
&\{Q_\alpha, Q_\beta\}= 4 P^- \delta_{\alpha\beta} + 2 Z^a
\gamma_{a\alpha\beta} + 2 Z^{a_1...a_4}
\gamma_{a_1...a_4\alpha\beta} \cr}}
In comparing \lcSUSY\ with \susyalg\ note that we have rescaled the
supercharges by numerical constants.  More important is the way the
central charges $Z^{\mu_1\mu_2}$ and $Z^{\mu_1...\mu_5}$ are handled.
Since the time in the light cone frame is $x^+$ we have set to zero the
charges with a component along $\mu=+$.  The charges with components
along $\mu=-$ are denoted by $Z^a$ and $Z^{abcd}$.  They
are activated by two branes and five branes which are stretched along
the longitudinal direction.  Clearly, to keep them finite we must
compactify this direction on a circle of radius $R$ and then they both
scale like $R$.

{}From this algebra we can distinguish three types of BPS states in the
infinite momentum frame (IMF).

\item{1.} Purely transverse membranes and fivebranes.  For these, half of the
SUSYs are preserved.  They are linear combinations of the kinematical
and dynamical generators.  The IMF energy of these states is
proportional to the square of the central charge $(Z^{ab})^2$ or
$(Z^{abcde})^2$.  This corresponds to the fact that if the
transverse directions are compactified, these excitations propagate as
particles.  For such states the IMF energy (for zero transverse
momentum) is proportional to the mass squared divided by the
longitudinal momentum $P^+$.  Thus, the corresponding brane tensions are
proportional to the central charge.  The authors of
\ref\dhn{B. de Wit, J. Hoppe and H. Nicolai, \np{305}{1988}{545}. B. de
Wit, M. Luscher and H. Nicolai, \np{320}{1989}{135}.},
who were the first to demonstrate that the matrix model describes
membranes, showed that membrane energies indeed scale like the inverse
longitudinal momentum in the matrix model.  This scaling is somewhat
implicit in their formalism, because they did not make the
identification of the longitudinal momentum with the rank of the
matrices which was proposed in \bfss.  It follows from the relation
between matrix commutators and Poisson brackets on the membrane volume.

\item{2.} Branes wrapped around the longitudinal direction.  These have
nonzero values of $Z^a$ or $Z^{abcd}$, with all other central charges
vanishing.  They preserve one quarter of the SUSYs.  All kinematical
generators are broken and half the dynamical ones are preserved.  The
energy of such a state is simply proportional to the central charge.
Furthermore it scales like a constant in the $P^+ \rightarrow\infty$
limit.  A wrapped object can carry momentum in the direction of the
circle on which it is wrapped only if it contains some internal
excitation which breaks translation invariance around the circle.  As
the momentum is scaled to infinity, the internal excitation energy also
goes up, so that the energy in the IMF does not scale like
$1/P^+$.\foot{An explicitly calculable example of this phenomenon arises
in first quantized string theory.  The energy $E$ of a state satisfies
$E^2= \half P_L^2 + \half P_R^2 +N_L+N_R$ where $P_L$ and $P_R$ are
given in terms of the momentum $P$ and the winding $L$ as $P_L= P+L$ and
$P_R=P-L$.  If $L$ in some direction is not zero and we let $P$ in this
direction go to infinity, the level matching condition $\half P_L^2 -
\half P_R^2 +N_L- N_R = 2PL +N_L-N_R=0$ implies that $|N_L- N_R|$ goes
to infinity like $|P|$.  Therefore, the ``mass'' square $m^2=E^2-P^2$ is
not a constant in this limit but grows linearly with $|P|$, so that the
IMF Hamiltonian $E-|P|$ goes to a constant.}  It is somewhat remarkable
that the lightcone SUSY algebra automatically takes this dynamical fact
into account.  This internal excitation is also the source of the
breaking of the extra half of the SUSY generators.

\item{3.} Branes wrapped around the longitudinal direction which
preserve half the supersymmetry.  As an example consider a brane with
nonzero values for both the two and the four brane charges, related by
the formula (for a particular choice of orientation of the solution)
$P^+ Z^{1234} = Z^{12}Z^{34}$.  This preserves half of the SUSYs, again a
combination of the kinematical and dynamical generators.  It is a
longitudinal five brane, with two orthogonal infinite stacks of two
branes embedded in it. If we insist that the four brane charge is
finite, then the product of two brane charges must scale like the 
longitudinal momentum.  

In terms of the fields in the low energy supergravity Lagrangian the
supercharges appear as integrals of total derivatives, and are
nonvanishing only on topologically nontrivial field configurations.  In
relating them to the matrix model \bfss\ our guiding philosophy is that
the analogous notion is that of commutators of infinite matrices which
have nonzero trace.  This idea, already hinted at in \bfss, is closely
related to an interpretation of nonvanishing central charges as
topological objects (integrals of total derivatives).

The plan of the rest of this paper is as follows: In the next section we
compute the commutator of supercharge densities in the matrix model and
show that it contains terms which may be interpreted as components of
the membrane and five brane charge densities.  We compare this to a
computation of membrane supercharge densities.  

In section $3$ we find
all static classical configurations of the matrix model which preserve
half of the supersymmetries.  Apart from the original membrane
discovered in \bfss, these are $p$-brane configurations with infinite
stacks of orthogonal membranes embedded in them.  The simplest of these
extends in four transverse directions.  We show that it satisfies a BPS
formula with nonzero values of the five brane charge, and may thus be
interpreted as a five brane wrapped around the longitudinal dimension.
It differs from the Berkooz-Douglas fivebrane, in that it carries
membrane charge as well.  The energy of the configuration scales in a
manner consistent with this interpretation.  Other configurations, with
extent in 6 and 8 transverse dimensions, do not seem to have a
conventional M theory interpretation.  However, it may be that they
correspond to D-brane excitations of string theory and only make sense
after compactification to 10 dimensions.  We show (in section 4) that
the Lagrangian for small fluctuations around these configurations indeed
reduces to the world volume Lagrangian of a D-brane.  We note however
that the tensions (energy densities) of these branes scale to infinity
in the large $N$ limit with eleven noncompact dimensions.  We have not
found a satisfactory interpretation of this result.  Finally, we exhibit
classical matrix configurations describing the longitudinal five brane
with no membrane charge.

In section $4$ we compute the small fluctuations around all of these
configurations and verify that they include the collective coordinates
implied by the $p$-brane interpretation.  We also show how enhanced
gauge symmetry arises when parallel two branes are brought together.

In the conclusions we point out that the matrix model supercharge
density algebra does not have a term corresponding to the purely
transverse components of the five brane charge density. 
We suggest that this may be a defect of the light cone gauge, related to
the fact that purely transverse D-branes cannot be constructed in light
cone gauge perturbative string theory.  We present
arguments which suggest that the transverse fivebrane cannot be a
classical solution of the matrix model.

As this paper was being written, a revised version of \princeton\
appeared, which also discusses the four brane charge and the self-dual
solution.

\newsec{\bf The Supercharge Density Algebra}

In this section we will construct the matrix analog of the supersymmetry
algebra.  The analog of integration over the membrane volume is taking
the trace of a matrix.  The SUSY generators are written as traces of
products of matrices, and it is natural to define the untraced products
to be the supercharge densities.  As with any density algebra, we are
free to add ``improvement terms'' to the densities, which do not affect
the traced charges, at least for a large class of configurations of the
system.  For the present computation we will content ourselves with a
minimal improvement which insures that the supercharge densities are
hermitian matrices.  This does not affect the supercharges for finite
$N$, and the improvement also vanishes in the smooth membrane
approximation to the infinite $N$ system.

We use the conventions for nine dimensional gamma matrices
\eqn\conv{\eqalign{
&\{\gamma^a,\gamma^b\}=2\delta^{ab} \cr
&\gamma^{ab}= \half [\gamma^a,\gamma^b]\cr}}
and the other $\gamma^{ab...}$ normalized similarly.
Recall the symmetry properties in the spinor indices
$\delta_{(\alpha\beta)}$, $\gamma^a_{(\alpha\beta)}$, $\gamma^{ab}
_{[\alpha\beta]}$, $\gamma^{abc} _{[\alpha\beta]}$,
$\gamma^{abcd}_{(\alpha\beta)}$.
We need the identity
\eqn\idef{I^b_{\alpha\beta\alpha'\beta'}= \gamma_{\beta\beta'}^a
\gamma_{\alpha\alpha'}^{ab} + \gamma_{\beta\beta'}^{ab}
\gamma_{\alpha\alpha'}^a + (\alpha \leftrightarrow \beta) =
2(\gamma^b_{\alpha'\beta'} \delta_{\alpha\beta} -
\gamma^b_{\alpha\beta}\delta_{\alpha'\beta'}).}

As in \bfss\ we study the Lagrangian
\eqn\lagran{\CL= \Tr  L}
where the Lagrangian density $L$ is the $N \times N$ matrix
\eqn\lagden{\eqalign{
&L= {1 \over 2R} (D_0 X^a)^2 + \theta^\alpha D_0 \theta^\alpha + {R
\over 4} [X^a,X^b]^2 + {iR \over 2} [\theta^\beta ,
[X^a,\theta^\alpha] ] \gamma^a_{\alpha\beta} \cr
&D_0 X^a = \partial_0 X^a - i [A_0,X^a] \cr
&D_0 X^a = \partial_0 \theta^\alpha- i [A_0,\theta^\alpha] . \cr}}
$\theta^\alpha$ ($\alpha=1,...,16$), $X^a$ ($a=1,...,10$) and $A_0$ are 
hermitian $N\times N$ matrices and $R$ is the radius of the longitudinal
direction.  The commutators in \lagden\ are 
commutators of these matrices.  The last term in $L$ differs from the
standard way of writing $L$ by an anticommutator of odd variables.  For
finite matrices this does not affect its trace, $\CL$.  It was added
here to make $L$ hermitian (we use the convention that for $u,v$ odd
Grassman numbers $(uv)^*=u^*v^*$).  Matrices have upper and lower $SU(N)$
indices and are multiplied according to 
\eqn\matrixmul{(AB)_i^j=A_i^k B_k^j.}

The Hamiltonian corresponding to this Lagrangian in the $A_0=0$ gauge is
\eqn\hamil{\CH=R \Tr  h}
with the hermitian Hamiltonian density
\eqn\hamden{h=\half P^2 - {1 \over 4} [X^a,X^b]^2 -{i \over 2}
[\theta^{\alpha},[X^b,\theta^{\beta}]] \gamma^b_{\alpha\beta}.} 
The Dirac brackets (DB) are
\eqn\piossonb{\eqalign{
&\big[X_i^{aj}, P_k^{bl} \big]_{DB}= \delta^{ab} \delta_i^l
\delta_k^j \cr 
&\big\{ \theta_i^{\alpha j}, \theta_k^{\beta l} \big\}_{DB} ={1\over
2} \delta^{\alpha \beta} \delta_i^l \delta_k^j. \cr}}
We note that in the computation which we will perform, no quantum
mechanical operator ordering ambiguities arise, so that the Dirac
bracket computation captures the full quantum operator algebra.  We have
retained the Dirac bracket nomenclature in order to avoid confusion
between commutators of quantum operators and commutators of matrices.

We define the supercharges
\eqn\defsup{\eqalign{&Q_\alpha = \Tr  q_\alpha \cr
&\tilde Q_\alpha = \Tr  \tilde q_\alpha \cr}}
where the densities $q_\alpha$ and $\tilde q_\alpha$ are $N \times N$
matrices 
\eqn\defq{\eqalign{
&q_{\alpha i}^j = \sqrt{R} \big\{P^a \gamma^a_{\alpha \alpha'} + {i
\over 2} [X^a,X^b]\gamma^{ab}_{\alpha \alpha'} , \theta^{\alpha'}
\big\}_i^j \cr  
& \tilde q_{\alpha i}^j = {2 \over \sqrt{R}}
\delta_{\alpha\alpha'}\theta_i^{\alpha'j} \cr}} 

We want to compute the Dirac brackets in a way which is sensitive to
traces of commutators.  In principle we should compute the Dirac bracket
of densities $\{ q_{i\alpha}^j , q_{k\beta}^l \}$. In this computation
we encounter terms which are odd under interchange of $\alpha$ and
$\beta$ and also under interchange of the matrix indices.  These are
analogs of Schwinger terms with odd derivatives of the delta function in
field theory charge density commutators.  The charges should be defined
as large $N$ limits of regularized traces of the densities.  As long as
we use the same regulator on both charges, these terms antisymmetric in
the spinor indices will not contribute to the Dirac bracket of the
charges.  As a consequence, when computing the DB of the dynamical SUSY
charge with itself, we can freely trace on one of the terms in the DB,
if we drop pieces of the answer antisymmetric in the spinor indices.
This simplifies the computation.  After some straightforward but
slightly tedious algebra we find:
\eqn\susyalg{\eqalign{
&\big\{\tilde q_{i\alpha}^ j, \tilde Q_\beta \big\}_{DB}=
{ 2 \over R } \delta_{\alpha \beta} \delta_i^j \cr
&\big\{q_{i\alpha}^j, \tilde Q_\beta \big\}_{DB}= 2\big( P^a
\gamma^a_{\alpha \beta} + {i \over 2} [X^a,X^b]\gamma^{ab}_{\alpha
\beta} \big)_i^j = 2P_i^{aj} \gamma^a_{\alpha \beta} +
2z_i^{abj}\gamma^{ab}_{\alpha \beta}\cr
&\big\{q_{i (\alpha}^ j, Q_{\beta)} \big\}_{DB}=
4 R (\half P^2 - {1 \over 4} [X^a,X^b]^2 )_i^j\delta_{\alpha \beta}
+ 2R \gamma^{abcd}_{\alpha \beta} \big(X^{[a} X^b X^c X^{d]}\big)_i^j  \cr 
& \qquad -2i R\gamma^b_{\alpha \beta} \big\{P^a,[X^a,X^b] \big\}_i^j
-i R [\theta^{\alpha'},[X^b,\theta^{\beta'}]]
I^b_{\alpha\beta\alpha'\beta'} \cr
&\qquad = 4 R h_i^j \delta_{\alpha\beta} + 2 z_i^{bj}
\gamma_{\alpha\beta}^b + 2 z_i^{abcdj} \gamma_{\alpha\beta}^{abcd} 
\cr}} 
where in the last anticommutator we symmetrized over $\alpha$ and
$\beta$.  We used the matrices
\eqn\defal{\eqalign{
&z^b= -i R\big\{P^a,[X^a,X^b] \big\}
-iR[\theta^{\alpha'}, [\theta^{\alpha'},X^b]]  \cr  
&z^{ab}=  {i \over 2} [X^a,X^b] \cr
&z^{abcd}= R X^{[a} X^b X^c X^{d]}\cr
}}

The matrices
\eqn\phidef{\left({\partial \CL \over A_0} \right)_j^j = \Phi_i^j= i
[P^a,X^a]_i^j + 2i (\theta^\alpha \theta^\alpha)_i^j} 
generate the $SU(N)$ algebra
\eqn\sualgebra{[\Phi_i^j,\Phi_k^l]_{DB}= i\big(\delta_i^l\Phi_k^j-
\delta_k^j \Phi_i^l \big).} 
The Gauss law constraint associated with the gauge $A_0=0$ is
$\Phi=0$.  Writing 
\eqn\newzb{z^b= - \{\Phi,X^b\} + i[\{X^b,P^a\},X^a]
+i\{\theta^{\alpha'},\{\theta^{\alpha'},X^b\}\}}
shows that for finite $N$, $\Tr z^b$ vanishes in the space of $SU(N)$
invariant states (satisfying Gauss law).

We interpret $\Tr z^{ab}$ as the two brane charge $ Z^{ab}$ and
$\Tr z^a$ and $\Tr z^{abcd}$ as the two charges of the wrapped branes
$Z^a$ and $Z^{abcd}$.  Note as a consistency check of this
interpretation the $R$ dependence of these charges.  $Z^{ab}$ is
independent of $R$ while $Z^{a}$ and $Z^{abcd}$ are proportional to
$R$.  All these charges are traces of commutators and therefore vanish
for finite $N$.  However, for infinite $N$ they can be activated.

de Wit, Hoppe and Nicolai \dhn, have computed the SUSY charge density
algebra on the membrane world volume in light cone supermembrane theory.
To compare our results with theirs, substitute $\theta
\rightarrow 2^{1/4} \theta$, $q \rightarrow 2^{1/4} q$ and $\tilde q
\rightarrow 2^{-1/4} \tilde q$ in our expressions.  Commutators of two
bosons $[A,B]$, commutators of a bosons and a fermion $[A,B]$ and
anticommutators of two fermions $\{A,B\}$ are all replaced by
$-i\epsilon^{rs} \partial_r A \partial_s B$; all other (anti)commutators
are simple.  Under this transcription our $z^{abcd}$ becomes zero.  The
longitudinal five brane charge vanishes in the membrane approximation to
the matrix model.  The authors of \dhn\ also have two terms which are
bilinear in $\theta$ which do not seem to follow from our computation.

\newsec{\bf BPS Branes in the Matrix Model}

As we have seen, light cone supersymmetry breaks up into two $16$
component supercharges.  Under the kinematical SUSY transformation the
fermionic coordinates transform as
\eqn\susytransa{\delta\theta = \tilde \epsilon}
The fermion transformation law under the dynamical SUSY transformations
is 
\eqn\susytransb{\delta\theta = \gamma_a P^a \epsilon + {i \over 2} [X^a ,
X^b]\gamma_{ab} \epsilon}
For static solutions, the only way to preserve half of the SUSY
generators, is to cancel the kinematical SUSY variation against the
dynamical one.  The kinematical SUSY transformation changes $\theta$ by
a multiple of the unit matrix, so we conclude that for static classical
BPS configurations of the matrix model, which preserve half of the
supersymmetries, the commutators of the $X^a$ must be multiples of the
identity:
\eqn\eom{[X^a , X^b] = i \CF^{ab} I}
where $I$ is the unit matrix in the $SU(N)$ space.  $\CF^{ab}$ is
antisymmetric, and can always be brought to canonical symplectic
(Jordan) form.  In \bfss\ it was shown that configurations of this type
indeed solve the static classical equations of the matrix model.  It is
clear that such configurations only exist in the large $N$ limit.

The static classical BPS states of the matrix model are thus
characterized by a set of orthogonal transverse two planes.  Apart from
the transverse membrane solution of \bfss, we have a four brane, a six
brane, and an eight brane.  (These names refer only to the transverse
dimensions.  We will see that the four brane should be interpreted as a
five brane wrapped around the longitudinal direction.)  We can verify
that the membrane solution is indeed the membrane solution of M theory
by examining the SUSY algebra.  The transverse two brane charge should
show up in the anticommutator of kinematical with dynamical SUSY
generators.  Indeed, as we have seen in the previous section, this
anticommutator contains, apart from the transverse momentum generator,
the trace of the commutator of two $X^a$ matrices, which is nonzero for
the membrane configuration.  The calculation of the membrane tension in
the appendix of \bfss\ can now be rederived as a BPS formula.  Note that
this anticommutator should also have contained the transverse five brane
charge, but it appears to be absent.

The transverse fourbrane configuration requires four of the $X^a$ to
satisfy the commutation relations of two pairs of canonical variables.
The uniqueness theorem for irreducible representations of the canonical
algebra tells us that we must represent this in the tensor product of
two Hilbert spaces.  Thus,
\eqn\cancomm{\eqalign{
&X^1 \propto q^1 \cr
& X^2 \propto p^1\cr
&X^3 \propto q^2 \cr
& X^4 \propto p^2 . }}
Here, the canonical variables are formal constructions for finite $N$
with commutators proportional to $N^{-\ha}$ (we will explain this
scaling below).  More generally, if $N=n_1n_2$, we can have one
commutator scale like $1 \over n_1 $ and the other like $1 \over n_2$.
Then, we take $n_1,n_2 \rightarrow \infty$.  Reducible representations,
corresponding to multiple branes, require a further tensor product with
a space in which all of the $X^a$ act as the unit matrix.  The four
brane clearly carries nonzero values of the transverse two brane charges
$Z^{12}$ and $Z^{34}$, which we constructed in the previous section.  We
interpret this in the following manner.  In the next section we will see
that smooth fluctuations around the fourbrane configuration can be
viewed as fields in the phase space defined by the canonical variables
$q^r , p^r$.  In this field space, the background configuration can be
viewed as representing two orthogonal infinite stacks of membranes.  The
$12$ stack is translation invariant in the $34$ directions, and {\it
vice versa}.

In addition, the fourbrane configuration has nonvanishing longitudinal
five brane charge $Z^{1234}$.  Remember that the corresponding charge
density vanishes in the membrane approximation to the matrix model.
Thus, the existence of this configuration, like that of the
supergraviton states, is a feature of the matrix model not shared by the
membrane Lagrangian.  The two kinds of charge are related via $P^+
Z^{1234} = Z^{12} Z^{34}$.  This is precisely the relation necessary to
ensure the BPS condition for the eleven dimensional SUSY algebra.  Thus,
we can interpret our transverse four brane as a special configuration of
a longitudinal five brane with stacks of two branes in it.

By contrast, the transverse six and eight branes do not seem to
correspond to things we expect to find in M theory.  M theory
compactified on a circle does contain a six brane, but it is a ``Kaluza
Klein monopole'' and would be expected to decouple in the noncompact
limit.  We know of no seven or eight branes in M theory, and nine branes
are supposed to correspond to ``ends of the world'' which carry gauge
dynamics.  What then are these configurations which we have found?

Some insight can be gained by calculating the tensions of the various
branes.  For a brane with $2t$ transverse dimensions, the full space on
which the matrices act is a tensor product of $t$ spaces of dimensions
$n_l$ ($l=1...t$).  The total dimension is $\prod_l n_l =N$, where
$P^+=N/R$ is the total longitudinal momentum of the system.  We take
$n_l \rightarrow \infty$ such that $n_l \sim N^{1 \over t}$.  For smooth
semiclassical branes, we want to take matrices whose commutators
approach Poisson brackets and are interpretable as functions on a $2t$
dimensional phase space.  Thus, for the $2t$ brane we want $[X^a , X^b]
\sim N^{-{1\over t}}$.  The trace on the full vector space approaches
$N$ times the integral over phase space.  Thus, the tension contribution
to the energy of a $2t$ brane scales like
\eqn\tension{ \delta E \propto {\rm tr}\ [X^a , X^b ]^2 \sim N^{1 -
{2\over t}}}
For $t =1$, the membrane, this is proportional to $N^{-1}$, the
appropriate scaling for the energy of a finite object in the infinite
momentum frame.  For $t=2$ the energy is constant.  This, as we argued
in the introduction, is the appropriate behavior for a brane with one
dimension wrapped 
around the longitudinal dimension of the infinite momentum frame.

For branes of arbitrary dimension wrapped around a single circle, we
expect to find states of a similar energy, corresponding to longitudinal
momentum carried by brane waves which are translationally invariant in
the transverse directions.  In this case, the energy of the state
should be interpreted as a transverse energy density.  All of these
static solutions of the matrix model correspond to branes wrapped around
cycles of a transverse torus, where the size of the torus is encoded in
the periodicities implicit in \eom\ and \cancomm.  Shifts of the
$X^a$ by a lattice vector, are gauge transformations.  The versions of
these constructions appropriate for infinite eleven dimensional space
time has the proportionality constants in \eom\ scaled to infinity.
All energies scale to infinity in this limit, with finite transverse
energy densities.  Thus, the brane with $4$ transverse dimensions has
precisely the energy density we would expect for a five brane wrapped
around the longitudinal direction.  This interpretation also fits nicely
with the BPS formula, for the solution carries both membrane and
fivebrane charges, satisfying the relation which preserves half of the
SUSYs.

By contrast, for $t > 2$ the brane's transverse energy density scales to
infinity in the IMF.  We have not been able to find an interpretation
for these BPS states in terms of the conventional M theory menagerie.
Perhaps some sense can be made of them after further compactification.
Indeed, the computation of the next section shows that their world
volume Lagrangians are (at the level of quadratic fluctuations),
precisely those of the corresponding Dirichlet branes of perturbative
string theory. 

To conclude this section we will construct a wrapped fivebrane carrying
no membrane charge.  To this end, we note that the
transformation of the fermionic coordinates under dynamical SUSYs,
reduces for static configurations to:
\eqn\deltheta{\delta\theta ={i \over 2} [X^a, X^b ]\gamma^{ab}\epsilon}
Half of the variations vanish, if we take only four nonzero $X^a$
which satisfy the condition
\eqn\selfdual{[X^a, X^b ] = \ha \epsilon_{abcd} [X^c, X^d ]}
(we raise and lower indices freely using a flat Euclidean metric) where
$\epsilon_{abcd}$ is the Levi-Civita symbol in the four indices.
There are no solutions of this equation for finite matrices.
Multiplying by $[X^a , X^b]$ and taking the trace, we recognize the
right hand side as our wrapped fivebrane charge, which vanishes by
cyclicity of the trace.  A large $N$ solution can be obtained by
choosing the $X^a$ to be the covariant derivatives in a self-dual
Yang-Mills potential: 
\eqn\indexon{X^a = {1\over i}{\partial \over \partial_{Q^a}} - A_a (Q)}
It was recognized long ago 
\ref\casher{T. Banks, A.Casher, \np{167}{1980}{215}.} 
that such configurations are formal solutions of the
large $N$ equations for static solutions of our matrix model\foot{In
this reference the solutions are described as constant classical
solutions of the four dimensional Yang Mills equations.  M. Douglas and
M. Li (private communication) have also investigated this solution in
the context of the matrix model of M theory, but with a somewhat
different interpretation.}.  

Clearly, the minimally charged fivebrane is one for which we take an
$SU(2)$ gauge group.  The space on which our large $N$ matrices act is
the tensor product of four representations of the canonical commutation
relations\foot{Here we mean ``truly infinite dimensional matrices,''
whose commutator has no inverse power of $N$ in it.} -- $(Q^a, P^a
(\equiv {1\over i}{\partial \over \partial_{Q^a}}))$ and a doublet
representation of $SU(2)$. This configuration has finite transverse 
energy density in the IMF, as befits a longitudinal fivebrane.

 In conventional Yang-Mills theory, the
self-dual configuration with minimal topological charge has five
collective coordinates.  However, in the present context, the
translational collective coordinates are {\it gauge transformations} of
the large $N$ gauge group, generated by the unitary operators $e^{i b_a
P^a}$, where the $P^a$ are the canonical variables.  
The unitary transformation $e^{i c_a Q^a}$ shifts the $X^a$ variables in
\indexon\ by an arbitrary continuous four vector.  As in \bfss \ we
interpret this as meaning that the configuration is compactified on a
four torus, but now of zero radius.  If we compactify the $Q^a$ space on
a torus, the $X^a$ matrices are gauge equivalent only to discrete shifts
of themselves and we
obtain a configuration of the matrix model compactified on the dual
torus.  This configuration has no scale size collective coordinate 
(as we explain below).   Thus, we obtain a unique longitudinal fivebrane
configuration of the matrix model.  The uncompactified limit of this
configuration is a zero scale size instanton on a torus of zero radius.

There is another, superficially quite different, solution of these
equations.  Let $[q^m , p^n] = {2\pi i \over N^{1/2}} \delta^{mn};\ m =
1,2$ be two pairs of formal finite $N$ canonical pairs such as we
employed in the construction of the four brane with two brane charge.
Define
\eqn\whatsit{\eqalign{
&X^1 = \ha [ (p^1 - p^2)\sigma^3 + (p^1 + p^2)] \cr
&X^2 = \ha [ (q^1 + q^2)\sigma^3 + (q^1 - q^2)] \cr
&X^3 = \ha [ (p^2 - p^1)\sigma^3 + (p^1 + p^2)] \cr
&X^4 = \ha [ (q^2 +q^1)\sigma^3 + (q^1 - q^2)] \cr}}
These expressions also define solutions of the BPS condition, with 
the same energy and charges as \indexon.  We have not found a matrix
model gauge transformation which maps one into the other.  A preliminary
analysis of the fluctuations around \whatsit, suggests that perhaps this
configuration can be separated to two disjoint objects.  We do not know
how to interpret this fact.

\subsec{Compactification on a Four Torus}

Another view of these configurations is obtained by compactifying the
system on a four torus\foot{This section was motivated by remarks of L.
Susskind, who first noticed that the wrapped five brane was an
instanton.}.  As argued in \refs{\bfss,\compact,\princeton}, the
relevant 
variables of the matrix model are then most elegantly packaged as the
dimensional reduction of ten dimensional SYM theory on the dual four
torus (though this description makes certain assumptions about how the
large $N$ limit is to be taken).  The original zero branes and the
strings connecting them are described by the SYM variables.  The
membrane wrapped around a two torus is a toron
\ref\thooft{G. 'tHooft, \np{138}{1978}{1}; \np{153}{1979}{141}.}.
In this description the two brane charge and the membrane are visible
already for finite $N$ but the charge is conserved modulo $N$.  Finally, 
a longitudinal fivebrane wrapped around the four torus
becomes a point object in this description -- an instanton.

In string theory, we can consider the point fivebrane sitting at some
transverse distance from the collection of zerobranes (which look like
transverse D4-branes in this description).  The relative transverse
position is described by Higgs fields in the fundamental representation
of the large $N$ gauge group.  If the matrix model of M theory is
complete, then this must be equivalent to some configuration of the
zerobranes themselves.  Indeed, at large $N$, the quantum fluctuations
of the Higgs degrees of freedom are negligible.  The fivebrane is
completely described by the classical ``image'' it makes in the $4+1$
dimensional SYM theory, which is well known to be an instanton
\ref\who{E.Witten, \np {460}{1996}{541}, hep-th/9511030; M.Douglas,
hep-th/9512077, hep-th/9604198.}.
We conclude that on a $4$ torus, the correct description of the wrapped
fivebrane is as an instanton solution of the $4+1$ dimensional gauge
theory on the dual torus (which of course is a soliton in this context).
On a finite torus, the scale of an instanton is not a collective
coordinate, but instead is determined.  The action is bounded below by
the topological charge, and this bound is achieved by the zero size
instanton.  Note that in the decompactification limit, the dual torus
shrinks to zero radius, so we would not expect to regain the collective
coordinate for the scale of the instanton.

This compactified picture also sheds some light on the fivebranes with
twobrane charge which we discussed above.  SYM theory on a torus has
toron configurations \thooft\ which carry topological charge equal to
the product of $Z_N$ fluxes in orthogonal planes.  Configurations whose
topological charge derives solely from $Z_N$ flux have twice as much
SUSY as instantons, and correspond to the wrapped fivebranes with
embedded two branes which we described above.

We can give a unified description of all the branes we discussed in the
framework of the compactified SYM theory.  We use $\pi_{2k-1}(U(N)) = Z$
for $N \ge k$, which is measured by $Q_k= \int \tr F^k$.  Then $Q_k$ is
a $2k$ brane charge; i.e.\ a gauge configuration with non-zero $Q_k$ is
a $2k$ brane.

We would like to emphasize that a comparison of the discussion of the
present subsection with that immediately preceding it shows once again
that all of the physics of the compactified theory is completely encoded
in the large $N$ matrix quantum mechanics with no additional degrees of
freedom.

\newsec{\bf Fluctuating Branes}

In this section we study the small fluctuations around our BPS
configurations.  For simplicity, we consider only the bosonic degrees of
freedom.  The fermions can easily be added.

We have shown that BPS solutions preserving $16$ linear combinations of
the kinematical and dynamical SUSYs satisfy.
\eqn\sol{[X^a,X^b] = i \CF^{ab} I }
where $\CF^{ab}$ are numbers and $I$ is the unit matrix, Using the
$SO(9)$ symmetry we can bring $C$ to a Jordan canonical form $\CF^{12}=
- \CF^{21} $, $\CF^{34}= - \CF^{43} $, $\CF^{56}= - \CF^{65} $,
$\CF^{78}= - \CF^{87} $ and all others vanish.

We replace the indices $a,b$ by $r,s=1,...,n$ and $I=n+1,...,9$ such
that $\CF^{IJ}=\CF^{rI}=0$ and again freely raise and lower indices
using the flat Euclidean metric.  We expand the matrix variables around
the classical solution $X^r=U_r$, $X^I=0$
\eqn\expansion{X^r=U_r + A_r .}
We define the ``field strength''
\eqn\deff{\eqalign{
&F_{0r}=-F_{r0}= \partial_0 A_r + i[U_r,A_0] \cr
&F_{rs}= -i[U_r,A_s] + i[U_s,A_r] \cr}}
which is gauge invariant under
\eqn\gaugetran{\eqalign{
&\delta A_0=\partial_0 \lambda \cr
&\delta A_r= -i[U_r,\lambda] .\cr}}
This gauge transformation is the linearized version of the $U(N)$ gauge
transformation
\eqn\ungauge{\eqalign{
&\delta A_0= \partial_0 \lambda + i [\lambda,A_0] \approx  \partial_0
\lambda\cr 
&\delta A_r = -i[U_r, \lambda] + i [\lambda, A_r] \approx -i[U_r,
\lambda] \cr
&\delta X^I= i[\lambda,X^I] \approx 0 .\cr}}
The field strength satisfies the ``Bianchi identity''
\eqn\bianchi{\partial_0F_{rs} -i[U_r,F_{s0}] -i[U_s,F_{0r}] =0.}
Both in verifying gauge invariance under \gaugetran\ and in checking
\bianchi\ one must use the Jacobi identity and the fact that $[U_r,U_s]$
is proportional to the unit matrix.

We expand the bosonic terms in the Lagrangian to quadratic order in
$X^I,A_0,A_r$ and we drop traces of commutators except $\Tr [U_r,U_s]=
iN\CF^{rs}$ 
\eqn\lagexp{\eqalign{
&\Tr~ \left( \half (D_0 X^I)^2 + \half (D_0 X^r)^2 + \half [X^r,X^I]^2 +
{1 \over 4} [X^r,X^s]^2 + {1 \over 4} [X^I,X^J]^2 \right) \cr
& \approx \Tr  \left( {1 \over 4} F_{0r}^2+ {1 \over 4} F_{r0}^2 - {1
\over 4} F_{rs}^2 +\half (\partial_0 X^I)^2 + \half [U_r,X^I]^2  -{1
\over 4} (\CF^{rs})^2 \right).\cr}} 

For the special case that only $\CF^{12} =-\CF^{21} \not= 0$ ($n=2$) we can
dualize the gauge field $A_0,A_r$ to a single gauge invariant field
$\phi$
\eqn\duala{\eqalign{
&F_{12}=\partial_0 \phi \cr
&F_{0r}= -i[U_s,\phi] \epsilon_{rs} .\cr}}
Under this substitution the equation of motion of $F$ is satisfied
trivially while the Bianchi identity of $F$ is ensured by the equation
of motion from the new Lagrangian
\eqn\newlag{\half \Tr  \left( (\partial_0 \phi)^2 + (\partial_0 X^I)^2
+ [U_r,X^I]^2 + [U_r,\phi]^2 - (\CF^{12})^2 \right) .}

Returning to the general case $r=1,...,n$, we can now view $U_r$ as
coordinates on an $n$ dimensional phase space.  Then, the matrices
$X^I,A_0,A_r$ can be replaced by functions on this phase space.  The
commutators with $U_r$ become derivatives
\eqn\commder{-i[U_r, \CO] = \partial_r \CO}
for any function on phase space $\CO$.  Note that the fact that the
$U_r$'s do not commute is consistent with the fact that derivatives
commute:
\eqn\consder{\partial_r\partial_s \CO -\partial_s\partial_r \CO= -
[U_r,[U_s,\CO]] + [U_s,[U_r,\CO]] = [\CO ,[U_r,U_s]] =0}
where we again used the Jacobi identity and the fact that $[U_r,U_s]$ is
proportional to the unit matrix.

This turns \lagexp\ to a $n$ dimensional Lorentz covariant Lagrangian
including a $U(1)$ gauge field and $9-n$ scalars.  This is the
Lagrangian describing D$n$-branes in the IIA string theory.  For the
special case of 2-branes, the dual variable $\phi$ corresponds to the
eleventh dimension.

Thus, we have verified that the small fluctuations around these BPS
solutions of the matrix model contain the collective coordinates of
branes of appropriate dimension, which justifies our name for these
objects.  It would be interesting to repeat these calculations for the 
longitudinal fivebranes with no membrane
charge.

We can also verify the existence of enhanced gauge symmetries when two
or more parallel membranes of the same charge are brought together.
In doing so it is important to remember that in the above calculations
we have been using the Poisson bracket approximation for the commutators
of matrices which are functions of the canonical variables which define
the membrane background.  When $M$ parallel membranes sit at the same
point in the transverse dimensions, the classical solution becomes the
above single membrane solution tensored with the unit matrix in an $M$
dimensional space.  This is an example of the block diagonal
construction of multi-extended-object states which was described in
\bfss\ .  When we consider small fluctuations around the classical
configuration, they are described by general hermitian matrices in this
$M$ dimensional space.  The commutator of the two coordinates in the
membrane direction is, in the Poisson bracket approximation,
\eqn\gauge{[X^1 , X^2] = \partial_1 A_2 - \partial_2 A_1 + i [[A_1,
A_2]],} 
where $A_r$ is defined as above, but is now a matrix valued field. The
double bracket symbol $[[A , B]]$ 
refers to commutators of $M \times M$
matrices.  It is easy to see that the electric field strength and gauge
transformation laws described above also generalize to the appropriate
$U(M)$ covariant formulae.  The terms involving transverse coordinates
become the covariant derivatives for Higgs fields in the adjoint of
$U(M)$.    Thus, in the Poisson bracket approximation, the multimembrane
system has a $U(M)$ enhanced gauge symmetry when the membrane positions
coincide. 

Note that in the full matrix model, the ultraviolet properties of this
gauge theory will be regularized by the noncommutative geometry of the
membrane volume.  However, its infrared behavior is unaffected.  Since
the gauge coupling is a relevant operator the theory evolves to strong
coupling in the infrared.  The moduli space of vacua of the theory is
the Cartan subalgebra of the Lie algebra of the gauge theory, modded out
by the Weyl group.  It is possible that the theory at the origin is a
free field theory with the Weyl group acting as a gauge symmetry (an
orbifold).  Alternatively, the theory there could be at a non-trivial
infrared fixed point of the renormalization group.  Either way, the
off-diagonal gauge bosons of the gauge group are not the right degrees
of freedom at long distance.  The matrix model tells us nothing new
about these possibilities.  It only shows us that the phenomenon of
enhanced gauge symmetry for coinciding branes occurs in $M$ theory, just
as it does for D-branes in perturbative string theory\foot{For membranes
in flat 11 dimensional space this gauge theory is at infinite coupling
and therefore there should not be any remnant of the off diagonal gauge
bosons (of course, they are important upon compactification).}.

\newsec{\bf Conclusions}

The SUSY algebra of the matrix model contains an enormous amount of
information and has led us to the discovery of a class of p-brane
solitons including the longitudinal fivebrane predicted
by M theory.  This makes it all the more striking that there is no
vestige of the transverse fivebrane charge in the matrix model.
This is incompatible with Lorentz invariance, so it is clear that there
is something missing in the rules for the matrix model which have been
elaborated up to this point.  
We
emphasize that this charge should have appeared in the ``easy''
anticommutator between the kinematical and dynamical SUSY generators.
Although we can imagine adding ``improvement terms'' to the SUSY
generators which generate terms with the right tensor structure, they
are somewhat arbitrary and have not as yet shed any light on the puzzle.

There are two logically separate issues:  
\item{1.}  Does the finite $N$ matrix model contain states which become
transverse fivebranes in the large $N$ limit? 
\item{2.}  What modifications of the matrix model rules must we make in
order to incorporate the fivebrane charge into the matrix model SUSY
algebra?  

\noindent
At the moment we have two sets of clues which appear to give somewhat
contradictory answers to the first of these questions.  The first arises
in the context of compactification.
 As currently understood, the rules for torus
compactification of the matrix model lead to large $N$ SYM theory on the
dual torus.  In the case of a transverse three torus, field theoretic
S-duality leads \princeton\ to a description of the wrapped five
brane in terms of a classical configuration of the dual Higgs fields
of the SYM theory.  This suggests that the matrix model contains the
transverse fivebrane, but that it is not a classical configuration of
the original matrix variables.  It seems extremely important to find a
more explicit construction of this configuration and to decompactify it.

On the other hand, there is one well known fact about perturbative
string theory which suggests that the puzzle of transverse five branes
may be an artifact of our enforced reliance on a light cone gauge
description of the matrix model.  {\it There are no transverse D-branes
in light cone gauge string theory.}  Transverse D-branes are not
longitudinally translation invariant and when boosted to the IMF, they
are no longer static objects\foot{ More formally, the Virasoro
condition, $\partial_s X^{-} = \partial_s X^a \partial_t X^a$ (where $s$
and $t$ are world sheet space and time coordinates) shows that if we
choose either Dirichlet or Neumann boundary conditions for the
transverse variables, the longitudinal coordinate always satisfies
Neumann boundary conditions.}.  Note that large transverse strings {\it
are} allowed in the light cone formalism, because strings are not
D-branes.  A parallel can be made between these observations and our
discovery of transverse membranes, but not five branes, in the matrix
model.  Membranes appear to be the elementary objects of the matrix
model, while fivebranes are expected to be D branes
\ref\andypaul{A. Strominger, \pl{383}{1996}{44}, hep-th/9512059;
P. Townsend, \pl{373}{1996}{68}, hep-th/9512062.}.
This comforting analogy suggests that the missing five brane charge
might be only a gauge artifact, but it fails to account for the
existence of the wrapped transverse fivebrane of \princeton\ .  

We are thus left with an unresolved puzzle.  The current formulation of
the matrix model may well contain dynamical fivebranes and admit a
covariant generalization.  Alternatively, a fully covariant,
nonperturbative formulation of M theory may have to introduce membranes
and five branes on an equal footing as elementary objects.  The
resolution of this conundrum is of the utmost importance.

\bigbreak\bigskip\bigskip

\centerline{\bf Acknowledgments}\nobreak
This work was supported in part by DOE grant DE-FG02-96ER40559.  We
thank J.Distler and M.Douglas for discussions.

\listrefs
\end